\documentstyle[11pt,paspconf,psfig]{article}
\def\edcomment#1{\iffalse\marginpar{\raggedright\sl#1\/}\else\relax\fi}
\reversemarginpar

\begin{document}
\title{Dating Starburst and Post-Starburst Galaxies}
\author{Rosa M. Gonz\'alez Delgado}
\affil{Instituto de Astrof\'\i sica de Andaluc\'\i a (CSIC). Apdo. 3004, 18080 Granada, Spain.}

\begin{abstract}

Stellar population models to predict the H Balmer and HeI lines in absorption are presented. Models are computed for burst and continuous star formation. It is found that the Balmer and HeI line profiles are sensitive to the age, except during the first 4 Myr of the evolution, when the equivalent widths of these lines 
are constant. The comparison of these lines with the corresponding nebular emission lines indicates that H$\delta$ and the higher-order terms of the Balmer series and HeI ($\lambda$3819, $\lambda$4026, $\lambda$4388 and $\lambda$4922) in absorption are very good age-indicators of starbursts. Models are applied to date the super-star cluster B in NGC 1569. 

\end{abstract}

\section{Introduction}

The techniques to date starbursts are based on their radiative properties which are determined by their massive stellar content. At ultraviolet wavelengths, the spectrum is dominated by absorption features many of them formed in the stellar wind of massive stars. The profiles of these lines are a function of the evolutionary state of the starburst (Leitherer, Robert, \& Heckman, 1995; Gonz\'alez Delgado, Leitherer, \& Heckman, 1997a). This technique has been successfully applied to age-date young starbursts ($\leq$10 Myr) (see Leitherer and Robert in this conference). The optical spectrum of starbursts is dominated by nebular emission lines formed in the surrounding interstellar medium of the starburst that is photoionized by radiation from the  massive stars.
This emission-line spectrum depends on the density and chemical composition of the gas, and on the radiation field from the ionizing stellar cluster; therefore, on the evolutionary state of the starburst (Garc\'\i a-Vargas, Bressan \& D\'\i az 1995). A test of consistency between these two techniques was performed in the prototypical starburst galaxy NGC 7714 (Gonz\'alez Delgado et al 1999). The age of the nuclear starburst derived from the wind resonance ultraviolet lines and the optical emission-line spectrum is 5 Myr old.

At the Balmer jump the spectra of starbursts can also show absorption features formed in the photospheres of O, B and A stars. The spectra of these stars are characterized by strong H and HeI absorption lines and with only very weak metallic lines. However, the detection of these stellar features at the optical wavelengths
in the spectra of starburst galaxies is difficult because H and HeI absorption
features are coincident with the nebular emission lines that mask the absorption. Even so, the higher-order terms of the Balmer series and some of the HeI lines are detected in absorption in many starburst galaxies or even in the spectra of giant HII regions (e.g. NGC 604, Terlevich et al 1996). These features can be seen in absorption because the strength of the Balmer series in emission decreases rapidly with decreasing wavelength, whereas the equivalent width of the stellar absorption lines is constant with wavelength. Since the strength of the Balmer and HeI absorption lines show a strong deppendency with the effective temperature and gravity, these lines are also an age indicator of starburst and post-starburst galaxies (D\'\i az 1988; Olofsson 1995).

In this contribution, evolutionary synthesis models are presented, to predict the profile of the H Balmer and HeI absorption lines of a single-metallicity stellar population up to 1 Gyr old. Models are used to predict the age of a super-star cluster in the starburst galaxy NGC 1569.

\section{Description of the models}

A stellar library of synthetic spectra which covers the main H Balmer 
(H$\beta$, H$\gamma$, H$\delta$, H8, H9, H10, H11, H12 
and H13) and HeI absorption lines (HeI $\lambda$4922, HeI $\lambda$4471,
HeI $\lambda$4388, HeI $\lambda$4144, HeI $\lambda$4121, HeI $\lambda$4026,  
HeI $\lambda$4009 and HeI $\lambda$3819) has been implemented in the evolutionary 
synthesis code Starburst99 (Leitherer et al 1999). Evolutionary models are computed for burst and continuous star formation up to 1 Gyr old and different assumptions about the stellar initial mass function. The stellar library is generated using a set of programs developed by Hubeny and 
colaborators (Hubeny 1988; Hubeny, Lanz \& Jeffery, 1995) 
in three stages. For T$_{eff}\geq$ 25000 K, the code TLUSTY is used to compute NLTE 
stellar atmosphere models. These models together with Kurucz (1993) LTE stellar 
atmosphere models (for T$_{eff}\leq$ 25000 K) are used as input to SYNSPEC, the 
program that solves the radiative transfer equation. Finally, the synthetic spectrum is obtained after performing the rotational (100 km s$^{-1}$ is assumed) and instrumental convolution. The final sampling of the spectra is 0.3 \AA. The metallicity is solar.

\section{Models results}

The Balmer and HeI line profiles are sensitive to the age (Figure 1), except 
during the first 4 Myr of the evolution, when the equivalent widths of these lines 
are constant. The equivalent widths of the Balmer lines range from 2 to 16 \AA\ and the HeI lines from 0.2 to 1.2 \AA.The strength of the lines is 
maximum when the cluster is a few hundred (for the Balmer lines) and a few ten 
(for the HeI lines) Myr old. In the continuous star formation scenario, the strength 
of the Balmer and HeI lines increases monotonically with time until 500 Myr and 100 Myr, respectively. However, the lines are weaker than in the burst  models due to the dilution of the Balmer and HeI lines by the contribution from very massive stars.

\begin{figure}

\psfig{file=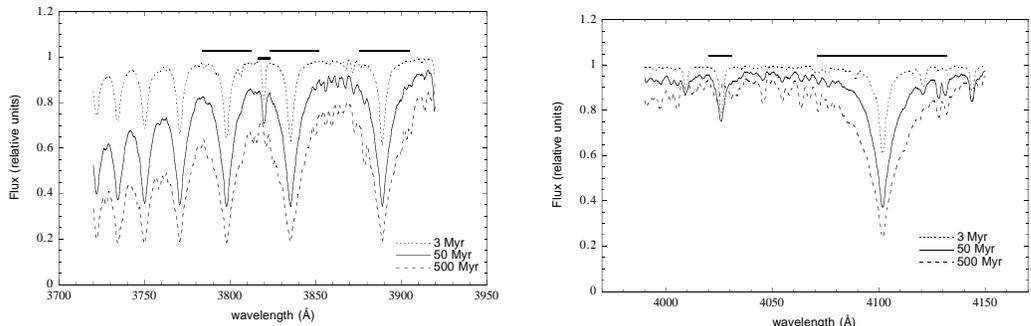,height=4.5cm}
\caption[fig]{Synthetic spectra from 3700 to 4200 \AA\ predicted for an instantaneous 
burst of 10$^6$ M$\odot$ formed following a Salpeter IMF between M$_{low}$= 1 M$\odot$ 
and M$_{up}$= 80 M$\odot$ at the age 3, 50 and 500 Myr.}
 
\end{figure}

As anticipated, the higher-terms of the Balmer series have equivalent widths similar to H$\beta$; therefore, this suggests that the higher-terms of the Balmer series are more useful to age-date starbursts than H$\beta$. To show which lines are better age indicators, the equivalent widths of the Balmer and HeI lines in emission are estimated.
Photoionization models using CLOUDY  (Ferland 1997) are computed to predict HeI $\lambda$4471/H$\beta$ assuming that the gas is spherically distributed 
around the ionizing cluster with a constant electron density. Figure 2 shows that H$\delta$ and the higher-order terms of the Balmer series and 
HeI are dominated by the stellar absorption component if an instantaneous burst is older than $\simeq$ 5 Myr. If the star formation proceeds continuously, after 30 Myr and 100 Myr, the strengths 
of the stellar absorptions are equal to those of the nebular emission lines H8 and H$\delta$, respectively. HeI $\lambda$4471 is very little affected by the absorption; however, the equivalent width of the emission line HeI $\lambda$3819 equals the stellar absorption line equivalent width at 20 Myr of continuous star formation, and after this time HeI $\lambda$3819 is dominated by the stellar absorption. 

\begin{figure}

\psfig{file=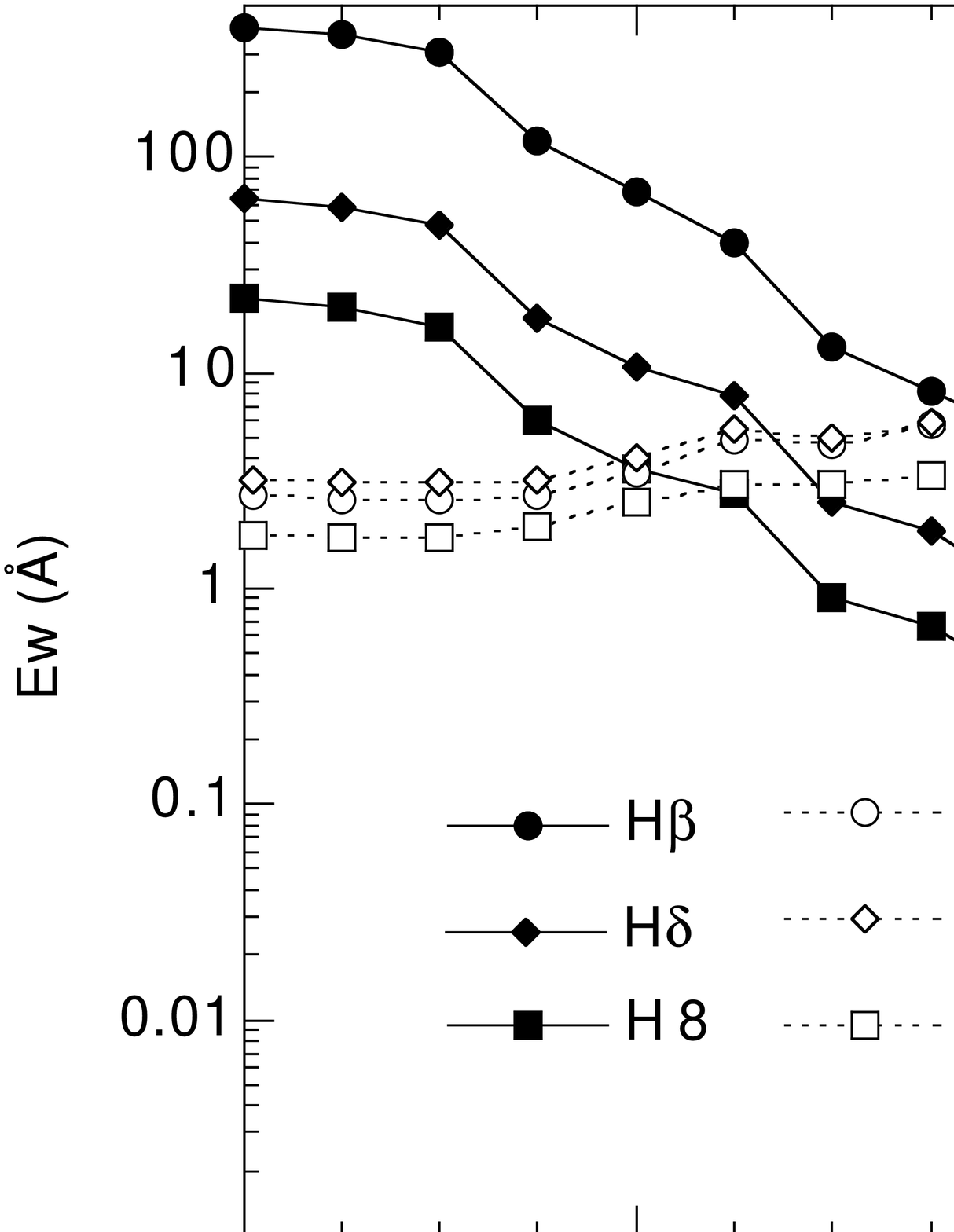,height=4.5cm}
\caption[fig]{Equivalent width of the Balmer (a) and HeI (b) lines for an instantaneous burst. The nebular
emission lines are plotted as full line and the absorption stellar lines as dotted lines}
 
\end{figure}

\section{Dating a super-star cluster in NGC 1569}

HST images of the starburst galaxy NGC 1569 suggest that the SSC B has an age of 15-300 Myr 
(O'Connell et al 1994) and ground-based optical spectra suggest an age of 10 Myr 
(Gonz\'alez Delgado et al 1997). The later result is based on the analysis of the 
optical spectral energy distribution. Since the Balmer lines are partially filled 
with nebular emission, the fitting has to be done based on the wings of the 
absorption features. Figure 3 plots the observed lines and the synthetic models for a burst of 10 and 50 Myr at Z$\odot$/4 metallicity (assuming Salpeter IMF, M$_{low}$=1 M$\odot$ and M$_{up}$=80 M$\odot$). The profiles 
indicate that the Balmer lines are more compatible with a burst of 10 Myr old than with
 the 50 Myr old. Ages older than 10 Myr produce profiles which are wider than the 
observed one. This comparison shows that this technique can discriminate well 
between a young and an intermediate age population.  

\begin{figure}

\psfig{file=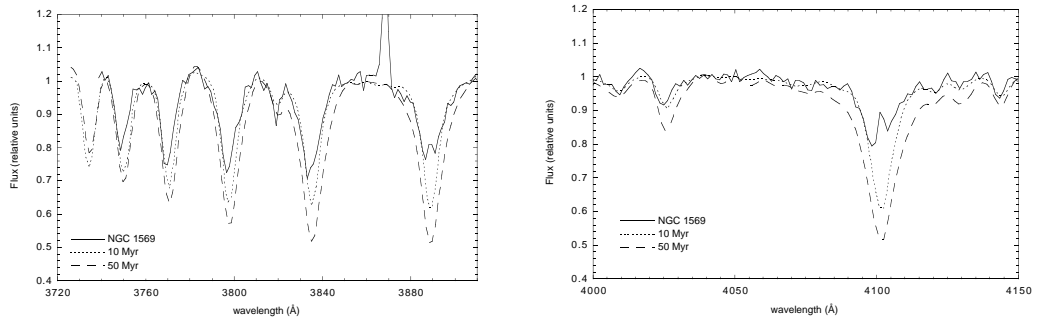,height=4.5cm}
\caption[fig]{Normalized optical spectrum of the SSC B in NGC 1569 (full line). The synthetic normalized spectra of an instantaneous burst 10 Myr (dotted line) and 50 Myr old (dashed line) formed following a Salpeter IMF at Z$\odot$/4 metallicity are ploted.}
 
\end{figure}

\end{document}